\documentclass[10pt,conference]{IEEEtran}
\usepackage{booktabs}   
\usepackage{array}      
\usepackage{caption}    

\AtBeginDocument{%
  }

\usepackage{amsmath,amsfonts, amssymb, amsthm}
\usepackage{algorithmic}
\usepackage{graphicx}
\usepackage{textcomp}
\usepackage{xcolor}
\usepackage{hyperref}
\usepackage{import} 
\usepackage[nameinlink]{cleveref}
\usepackage{xspace}
\usepackage{framed}
\usepackage{subfigure} 
\usepackage{listings}
\usepackage{soul} 
\usepackage{wasysym}
\usepackage{booktabs}
\usepackage{tabularx}
\usepackage{multirow}
\usepackage{pifont}
\usepackage{stmaryrd} 
\usepackage{svg}
\usepackage{balance}
\usepackage{dsfont}
\usepackage{microtype}

\usepackage{ifthen} 

\newboolean{showcomments}
\setboolean{showcomments}{true} 
\ifthenelse{\boolean{showcomments}}{
  \newcommand{\nbc}[3]{
    {\textcolor{#3}{\small{\bfseries{#1:\ }}\textit{#2}}}}
}{
  \newcommand{\nbc}[3]{}
}
\newcommand{\passat}[1]{\textrm{pass@#1}\xspace}

\newcommand{\numfixed}{\textrm{\#fixed}\xspace}

\newcommand{\ca}{\hbox{\emph{ca.}}\xspace} 

\newcommand{\etal}{\hbox{\emph{et al.}}\xspace}
\newcommand{\eg}{\hbox{\emph{e.g.}}\xspace}
\newcommand{\ie}{\hbox{\emph{i.e.}}\xspace}
\newcommand{\wrt}{\hbox{\emph{w.r.t.}}\xspace}
\newcommand{\etc}{\hbox{\emph{etc.}}\xspace}

\graphicspath{ {./pictures/} }

\definecolor{gray}{RGB}{215,215,215}
\sethlcolor{gray}

\DeclareRobustCommand{\hlyellow}[1]{{\sethlcolor{yellow}\hl{#1}}}

\theoremstyle{definition}
\newtheorem{definition}{Definition}[section] 
\newcommand{\projName}{\textsc{Maniple}\xspace}
\theoremstyle{finding}
\newtheorem{finding}{Finding}

\newcommand{\singleFunctionDataset}{\textsc{BGP314}\xspace}
\newcommand{\BGPSetA}{\textsc{BGP157Ply1}\xspace}
\newcommand{\BGPSetB}{\textsc{BGP157Ply2}\xspace}
\newcommand{\preliminaryStudyDataset}{\textsc{BGP32}\xspace}
\newcommand{\changes}[1]{%
	#1\xspace
}
\begin{document}

\title{The Fact Selection Problem\\ in LLM-Based Program Repair}



\author{\IEEEauthorblockN{Nikhil Parasaram}
\IEEEauthorblockA{\textit{University College London} \\
nikhil.parasaram.19@ucl.ac.uk}
\and
\IEEEauthorblockN{Huijie Yan${}^\dagger$}
\IEEEauthorblockA{\textit{University College London} \\
huijie.yan.20@ucl.ac.uk}
\and
\IEEEauthorblockN{Boyu Yang${}^\dagger$}
\IEEEauthorblockA{\textit{University College London} \\
boyu.yang.21@ucl.ac.uk}
\and
\IEEEauthorblockN{Zineb Flahy}
\IEEEauthorblockA{\textit{University College London} \\
zineb.flahy.21@ucl.ac.uk}
\and
\IEEEauthorblockN{Abriele Qudsi}
\IEEEauthorblockA{\textit{University College London} \\
abriele.qudsi.21@ucl.ac.uk}
\and
\IEEEauthorblockN{Damian Ziaber}
\IEEEauthorblockA{\textit{University College London} \\
damian.ziaber.21@ucl.ac.uk}
\and
\IEEEauthorblockN{Earl T. Barr}
\IEEEauthorblockA{\textit{University College London} \\
e.barr@ucl.ac.uk}
\and
\IEEEauthorblockN{Sergey Mechtaev}
\IEEEauthorblockA{\textit{Peking University} \\
mechtaev@pku.edu.cn}

}

\maketitle

\begin{abstract}
  Recent research has shown that incorporating bug-related facts, such as stack traces and GitHub issues, into prompts enhances the bug-fixing capabilities of large language models (LLMs). Considering the ever-increasing context window of these models, a critical question arises: what and how many facts should be included in prompts to maximise the chance of correctly fixing bugs? To answer this question, we conducted a large-scale study, employing over 19K prompts featuring various combinations of seven diverse facts to rectify 314 bugs from open-source Python projects within the BugsInPy benchmark. Our findings revealed that each fact, ranging from simple syntactic details like code context to semantic information previously unexplored in the context of LLMs such as angelic values, is beneficial. Specifically, each fact aids in fixing some bugs that would remain unresolved or only be fixed with a low success rate without it. Importantly, we discovered that the effectiveness of program repair prompts is non-monotonic over the number of used facts; using too many facts leads to subpar outcomes. These insights led us to define the fact selection problem: determining the optimal set of facts for inclusion in a prompt to maximise LLM's performance on a given task instance. We found that there is no one-size-fits-all set of facts for bug repair. Therefore, we developed a basic statistical model, named \projName, which selects facts specific to a given bug to include in the prompt. This model significantly surpasses the performance of the best generic fact set. To underscore the significance of the fact selection problem, we benchmarked \projName against the state-of-the-art zero-shot, non-conversational LLM-based bug repair methods. On our testing dataset of 157 bugs, \projName repairs 88 bugs, 17\% above the best configuration.
\end{abstract}



\def\thefootnote{${}^\dagger$}\footnotetext{These authors contributed equally to this work.}\def\thefootnote{\arabic{footnote}}

\section{Introduction}
\label{sec:intro}

When debugging and fixing software bugs, developers seek bug-related information from a diverse array of sources. Such sources include the buggy code's context, documentation, error messages, outputs from program analysis, etc. Individual pieces of this information, which following recent work we refer to as \emph{facts}~\cite{ahmed2023improving}, have been demonstrated by previous studies to enhance LLMs' bug-fixing efficacy when incorporated into the prompts~\cite{xia2023revisiting,fakhoury2023towards,xia2023keep}. Given the ever-increasing context window of cutting-edge LLMs, a critical question emerges: ``Which specific facts, and in what quantity, should be integrated into the prompt to optimise the chance of correctly fixing a bug?''

This work is a systematic effort to investigate how to construct effective prompts for LLM-based automated program repair (APR) by composing facts extracted from the buggy program and external sources. \changes{We identified seven facts: those individually studied in the context of APR by previous work, such as the buggy code's context~\cite{prenner2024out, xia2023revisiting, chen2024large}, GitHub issues~\cite{fakhoury2023towards}, and stack traces~\cite{Keller2024AI}; Angelic values, a semantic fact previously unexplored in the context of LLM-based APR, but that has been successfully used for debugging~\cite{chandra2011angelic} and repair~\cite{mechtaev2016angelix}; and those chosen based on our intution as developers. Our study was conducted on 314 bugs in open source Python projects from the BugsInPy~\cite{widyasari2020bugsinpy} benchmark.}

Our first experiment aims to confirm the utility of the considered facts. Specifically, for each fact, if the potential inclusion of this fact in APR prompts helps to repair some additional bugs, or increases the probability of fixing some bugs. To answer this question, we constructed over 19K prompts tasked to repair the buggy function and containing different subsets of the seven facts for the 314 bugs. Then, we queried an LLM to generate patches, and evaluated the patches using the provided test suites. \changes{\Cref{finding:utility} confirms the utility of each fact, that is each fact helped repair at least one bug that was not repaired by any prompt without this fact. Moreover, all the facts have statistically significant positive impact on the probability of fixing a bug in a single attempt.}

Given the utility of each fact, it is tempting to assume that adding more facts always enhances LLM's performance. \changes{Contrary to this intuition, \Cref{finding:monotonic} reveals that APR prompts are non-monotonic over facts: adding more facts may degrade LLM's performance.} An experiment involving 157 bugs showed that prompts incorporating all available facts resulted in 12 fewer bug fixes and exhibited an $8.2\%$ lower probability of repairing a bug within a single attempt compared to the most effective subset. \changes{This finding is non-obvious, because each fact may contain crucial information for fixing the bug, and although previous research showed that LLMs do not robustly make use of information in long input contexts~\cite{liu2023lost}, and their performance  dramatically decreases when irrelevant information is included in a prompt~\cite{shi2023large}, the trade-off between utility of information and the ability of LLM to process long contexts has not investigated.}

\begin{figure}[t!]
 \centering
 \begin{lstlisting}[escapechar=?,breaklines=true]
Please fix the buggy function provided below and output a corrected version.
?\hl{<.. CoT instructions are omitted ...>}?

## The source code of the buggy function
```python
# this is the buggy function you need to fix
def read_json(
?\hl{<... a part of code is omitted ...>}?
    return result
```
## A test function that the buggy function fails
```python
def test_readjson_unicode(monkeypatch):
?\hl{<... a part of code is omitted ...>}?
        tm.assert_frame_equal(result, expected)
```
### The error message from the failing test
```text
monkeypatch = <_pytest.monkeypatch.MonkeyPatch object at 0x7f567d325d00>
?\hl{<... a part of message is omitted ...>}?
pandas/_libs/testing.pyx:174: AssertionError
```
## Runtime values and types of variables inside the buggy function
compression, value: `'infer'`, type: `str`
?\hl{<... some variables are omitted ...>}?
lines, value: `False`, type: `bool`

## Expected values and types of variables during the failing test execution
path_or_buf, expected value: `'/tmp/tmphu0tx4qstest.json'`, type: `str`
?\hl{<... some variables are omitted ...>}?
result, expected value: ?\hl{<...omitted...>}?, type: `DataFrame`

## A GitHub issue for this bug
```text
Code Sample, a copy-pastable example if possible
?\hl{<... a part of text is omitted ...>}?
However, ?\hlyellow{when read\_json() is called without encoding parameter}?, it calls built-in open() method to open a file and open() uses return value of locale.getpreferredencoding() to determine the encoding?\hlyellow{which can be something not utf-8 }?
```
 \end{lstlisting}
 \vspace{-3mm}
 \caption{A simplified APR prompt incorporating various facts for fixing pandas:128. \hl{...} shows information omitted for brevity. \hlyellow{...} shows a part of the GitHub issue that is essential to correctly fix the bug. Too much information in the prompt ``distracts'' the LLM from the relevant part of the issue description, significantly reducing \passat{1} and the correctness rate.\label{fig:aprprompt}}
 \vspace{-3mm}
\end{figure}

The non-monotonicity of APR prompts and, simultaneously, the utility of each fact led us to define \emph{the fact selection problem}: determining the optimal set of facts for inclusion in prompts to maximise LLM's performance on given tasks. It can be viewed as a variant of feature selection of classical machine learning for LLM prompt engineering. We consider two instances of fact selection: \emph{universal fact selection} when the selected facts do not depend on a specific task instance, \ie the bug, and \emph{bug-tailored fact selection} when the fact set is bug-specific.

If there was a universal set of facts that is effective for all bugs, it would significantly simplify the development of LLM-based APR tools. However, our experiments showed that universal fact selection is suboptimal compared to bug-tailored fact selection. \changes{Specifically, \Cref{finding:universal} identified that there is no single subset of the seven considered facts that is sufficiently effective across all bugs in the dataset.} Meanwhile, enumerating sets of facts via \eg greedy strategies while repairing each bug might be impractical, because of the high cost of LLM queries and the necessity to generate multiple responses due to LLMs' nondeterminism. As a practical compromise, we trained a statistical model that we call \projName. It is designed to select facts contingent upon the features of a specific bug. \changes{Empirical evidence shows that \projName significantly outperforms a universal fact selection strategy, fixing $11$ more bugs than a generic set of facts that exhibited the optimal performance on the training data.}

We benchmarked \projName against state-of-the-art zero-shot non-conversational LLM-based APR techniques. On our testing set, \projName repaired 17\% more bugs, highlighting the practical impact of the fact selection problem.

The contributions of this work are:
\changes{
\begin{itemize}
\item A large scale systematic study of a diverse set of bug-pertinent facts for APR prompts.
\item An empirical evidence of the utility of each considered fact for program repair, including angelic values, previously unexplored in the context of LLMs.
\item An empirical evidence of the non-monotonicity of APR prompts over bug-related facts, \ie adding more bug-related facts into APR prompts may degrade LLM’s bug-fixing performance.
\item A motivation and introduction of the fact selection problem for LLM-based APR.
\item \projName, a bug-tailored fact selection model that, for a given bug, chooses a set of fact to construct an effective APR prompt; This model significantly outperforms previous related techniques.
\end{itemize}
}
All code, scripts, and data necessary to reproduce this work are available at \url{https://github.com/PyRepair/maniple}. 


\begin{figure}[t]
\vspace{-2mm}

 \centering
  \includesvg[inkscapelatex=false, width =\linewidth]{pictures/non-monotonic}
  \caption{Comparison of \passat{1} (the vertical axis) for around 10K prompts incorporating subsets of the seven considered facts (the horizontal axis) computed over 15 responses for repairing 157 Python bugs within the BugsInPy benchmark. Zero facts corresponds to the prompt containing only the buggy function without any additional information about the bug. The graph plots the average \passat{1} score (dashed lines) and the maximum \passat{1} score (solid lines) across all bugs. This graph clearly shows the non-monotonic nature of APR prompts over facts \changes{for GPT-3.5 and Llama3-70B.}}
  \label{fig:non_monotonic}
  \vspace{-2mm}

\end{figure}

\section{Motivating example}
\label{sec:motivating}

To illustrate the importance of the fact selection problem, we consider the bug pandas:128~\cite{pandas128bug} in the Pandas data analysis library within the BugsInPy benchmark. This bug arises due to incorrectly handling the default encoding in the Pandas' \lstinline{read_json} function. The developer patch for this bug involves setting the encoding to UTF-8 when it is not specified by adding the following lines to the buggy function:
\begin{lstlisting}[escapechar=?,breaklines=true]
+      if encoding is None:
+        encoding = "utf-8"
\end{lstlisting}

When GPT-3.5-Turbo~\cite{chatgpt} was prompted to rectify the bug solely based on the source code of the buggy function, it did not successfully address the issue in any of 15 \changes{responses}. A likely explanation of this failure is that the function itself does not contain any inherently incorrect code, so fixing this bug requires external information.

Drawing upon existing literature on LLM-based APR and relying on our intuition as developers, we assembled a diverse set of bug-related facts to incorporate into the prompt. These include the buggy function's context, the failing test case, the error message, the runtime values of local variables, their angelic values (values that, when taken by program variables during test execution, result in successful passage of the test), and the GitHub issue description. \Cref{fig:aprprompt} gives a simplified representation of the resulting prompt. Adding these facts enabled the LLM to generate a plausible patch, \ie a patch that passes the tests, in four out of 15 \changes{responses}. However, only two of these patches were correct. The other two hard-coded UTF-8 as the only encoding instead of the default one. The causes of failures to fix the bug included ``forgetting'' to change the function despite correct chain-of-thought reasoning~\cite{wei2022chain}, or hard-coding the encoding inconsistently.


Interestingly, when we removed all facts but the code of the buggy function, the runtime variable values and the GitHub issue, it significantly raised the success rate. Specifically, the LLM generated plausible patches in 12 out of 15 \changes{responses}, and 11 out of these 12 were correct. A similar high success rate was demonstrated by the prompt with only the buggy function and the GitHub issue. We posit that this is because redundant or irrelevant information in the original prompt ``distracted'' the LLM from critical details in the GitHub issue (highlighted in \Cref{fig:aprprompt}) necessary to repair the bug~\cite{shi2023large}.

We refer to this phenomenon, that adding more facts may degrade LLM's performance, as the \emph{non-monotonicity} of prompts over facts. To provide stronger empirical evidence, we conducted a large-scale experiment with 19K prompts containing various subsets of seven facts on 314 bugs in Python projects. \Cref{fig:non_monotonic} shows how \passat{1}, which estimates the probability of generating a plausible patch in a single trial, depends on the number of included facts. \changes{The graph shows two types of lines: the solid line corresponds to the scenario when we select the most effective combination of facts for each bug. The dashed line indicates the performance of an average fact set. It is evident that both these functions are non-monotonic for both GPT-3.5 and Llama3-70B. This phenomenon of non-monotonicity and its impact on prompt performance are discussed in more detail in \Cref{sec:eval:monotonicity}.}

Apart from the non-monotonicity of prompts, we also discovered that each of the considered facts helps to fix some bugs that cannot be fixed without it. These observations motivated us to formulate the \emph{fact selection problem}, the problem of selecting facts for a given bug to maximise the chance of repairing it and propose a model \projName that selects effective facts based on features of a given bug.

\section{Study Design}
\label{sec:setup}

This section discusses the experimental setup, the facts chosen for our study, and how they are represented in prompts.

\subsection{Experimental Setup}

We denote the set of bugs as $B$ and the set of all bug-relevant facts as $\mathbb{F}$. Since we aim to investigate how using various facts impacts the success of APR, we define the set of all jobs as $\mathbb{J} = B \times 2^{\mathbb{F}}$, which pairs bugs with sets of facts.

\paragraph*{Zero-shot Prompting for APR} A advantage of large language models, such as ChatGPT~\cite{chatgpt}, is they can be adapted to a downstream task without retraining via \emph{prompt engineering}~\cite{weng2023prompt}. A prompt refers to the input or instruction given to the model to elicit a response. Prompting typically takes either the form of \emph{zero-shot}, \ie directly providing the model a task's input, or \emph{few-shot}~\cite{brown2020language}, where the model is provided with a few examples. In this work, we investigate a zero-shot APR approach. Although the few-shot approach is promising, it requires finding high-quality examples~\cite{liu2021makes}, which we leave for future work.


\paragraph*{Function-granular Perfect Fault Localisation} APR tools repair bugs by first localising suspicious locations. For an objective evaluation of APR tools, Liu~\etal~\cite{liu2019you} argues for the use of \emph{perfect fault localisation (PFL)}, that is, when the buggy locations are known to the tool. PFL can resolve to difference granularity levels: notably, the line or the function. Liu~\etal argues that fault localisation tools do not offer the same accuracy in identifying faulty locations at different granularities, ``making function-level granularity appealing for limiting unnecessary \changes{responses} on fault positive locations''. We found that, in BugsInPy~\cite{widyasari2020bugsinpy}, the average Ochiai~\cite{abreu2007accuracy} rank of the buggy line is 2502 and the buggy function is 22 across the entire codebase, making it much more likely to localise the buggy function than the buggy line. Apart from that, although 70\% of the bugs in BugsInPy require modifying only a single function, 65\% of them modify multiple lines within this function. Localising a bug to multiple lines is harder than targeting individual lines. Meanwhile, cutting-edge models, like GPT-3.5-Turbo, effectively fix bugs even without specifying the exact lines, \ie when only the buggy function is provided.  \changes{Consequently, and in contrast to some previous studies~\cite{lutellier2020coconut,chen2019sequencer,xia2023revisiting}, this paper adopts function-granular PFL as the standard approach for evaluating APR tools. We implement this by providing the buggy function as context in the prompts for the LLM, enabling it to focus on fixing issues within the specified function.}

\begin{figure*}[t]
\vspace{-2mm}

  \includesvg[inkscapelatex=false, width =\linewidth]{pictures/facts}  
\caption{This work uses \changes{seven} facts (dark rectangles) across three categories for constructing program repair prompts. Each fact is composed of \changes{related pieces of information}. Each prompt contains the buggy function to be repaired; the facts can be included based on the employed fact selection strategy.\label{fig:facts}}
\vspace{-2mm}
\end{figure*}

\paragraph*{Python Bug Benchmark} \label{sec:back:bench} APR tools are typically compared on datasets of bugs extracted from real world projects, such as Defects4J~\cite{just2014defects4j} for Java and BugsInPy~\cite{widyasari2020bugsinpy} for Python. In this work, we use BugsInPy because of Python's ever-increasing importance and popularity. BugsInPy contains 501 bugs from 17 popular Python projects such as Pandas~\cite{pandasURL} and Matplotlib~\cite{matplotlibURL}. Among them, we selected a subset of 314 bugs, which we refer to as \textbf{\singleFunctionDataset}, that require modifications within a single function, due to our PFL approach, and that we were able to reproduce. To investigate APR performance on various classes of defects, we consider three parts of \singleFunctionDataset:
\begin{itemize}
\item \changes{
\textbf{\BGPSetA}: This dataset comprises 157 bugs, which have been uniformly selected from \singleFunctionDataset for the purpose of training and analysis.}
\item \changes{ \textbf{\BGPSetB}: Consisting of 157 bugs, this dataset is the complement of \BGPSetA in \singleFunctionDataset. Used for evaluating fact selection strategies.}
\item \textbf{\preliminaryStudyDataset}: A subset of 32 bugs uniformly sampled from \BGPSetA, intended for preliminary studies on finding parameters, such as determining the right response count to reduce the variance.

\end{itemize}

\paragraph*{LLM Nondeterminism}\label{sec:background:nondet} 
Nondeterminism in LLMs leads to varying outcomes between \changes{responses}, which poses a challenge for analysing results~\cite{ouyang2023llm}. To alleviate it, we use the \passat{k} measure, which represents the probability that at least one query out of k succeeds at solving a problem. Previous work~\cite{chen2021evaluating} recommends estimating \passat{k} as
\begin{equation}
\label{eq:pass@k}
    \passat{k}(\mathit{LLM}(Q)) \triangleq \mathbb{E}_{Q} \left[ 1 - \frac{{\binom{n-C}{k}}}{{\binom{n}{k}}} \right]
\end{equation}
where $\mathbb{E}_{J}$ denotes expectation over the set of LLM responses to the set of queries (prompts) $Q$, $n$ is the number of responses obtained from the LLM, where $n > k$ and $C$ is the number of successes found in the $n$ responses. \changes{Our task is program repair, so we deem a response successful if the extracted patch satisfies a correctness criterion, which we approximate with passing a test suite}. 
A pilot study using \preliminaryStudyDataset reveals that when $n=15$ and $k=1$, \passat{k} exhibits the average standard deviation of \ca 0.04, and that further increasing $n$ only marginally decreases standard deviation (Appendix A).

For generate-and-validate APR~\cite{qi2015analysis} that iteratively generates and tests patches until it finds one that passes, a commonly used measure is the number of bugs for which at least one patch passes the tests among LLM's responses:
\begin{equation}
\label{eq:numfixed}
    \numfixed(\mathit{LLM}(J)) \triangleq \big|\{b \mid j \in J,  C_j > 0\}\big|
\end{equation}
where $j = (b, F)$ is a prompt, and $C_j$ is the number of responses that pass the tests for the prompt $j$.

\paragraph*{Test-overfitting in Program Repair} APR techniques repair bugs \wrt correctness criteria, such as tests or formal specification. Since tests do not fully capture the intended behaviour, automatically generated patches based on tests may be incorrect~\cite{smith2015cure}. Thus, the APR literature distinguishes between \emph{plausible} patches, patches that pass the tests, and \emph{correct} patches, patches that satisfy the intended requirements. Since manually labelling a large number of patches is resource-intensive and error-prone, most analyses of the fact selection problem with \passat{k} and \numfixed in this paper count plausible patches as successes.  We only label correct patches when comparing our tool with other APR techniques in \Cref{sec:maniple_eval}.

In our experiments, we utilised the latest version of GPT-3.5-Turbo, specifically gpt-3.5-turbo-0125, which features a 16K context window. \changes{For our studies, GPT-3.5-Turbo was run with $n=15$ responses on \BGPSetA, \BGPSetB, and with $n=30$ for parameter exploration on \preliminaryStudyDataset. Additionally, we used llama3-70B, which offers an 8K context window and was run exclusively on \BGPSetA. As of March 2024, the cost of reproducing the experiments detailed in this paper, using the OpenAI API~\cite{openaiAPI} for GPT-3.5 and DeepInfra~\cite{DeepInfraAPI} for llama3-70B, is estimated at \$479 + \$269.}

\subsection{Bug-Related Facts}
\label{sec:facts}


The considered facts were collected from previous LLM-based APR research, previous non-LLM-based APR literature, and our intuition as developers. \changes{We conducted a pilot study on BGP32 to validate fact utility.}  In total, we collected 14 \changes{pieces of information, but since many of them are related, we grouped them into seven facts, which we refer to as $\mathbb{F}$}. These seven facts are divided into three categories: static, dynamic, and external, as shown in \Cref{fig:facts}.


\paragraph*{Buggy Class (1.1)} The declaration of a class containing the buggy function provides a broader context and dependencies. A class docstring offers insight into the overall purpose and functionality of the class.

\paragraph*{Used Method Signatures (1.2)} Considering methods used within the buggy function, as shown by Chen~\etal~\cite{chen2024large}, allows for the analysis of dependencies and potential side effects that might contribute to the incorrect behaviour.

\paragraph*{Failing Test (1.3)} The code of a failing test, as shown by Xia~\etal~\cite{xia2023keep} provides useful context for repairing a buggy function as it specifically highlights the conditions under which the program fails.

\paragraph*{Error Information (2.1)} Previous approaches showed that using error messages~\cite{xia2023keep} and stack traces~\cite{Keller2024AI} improves LLM's bug-fixing performance.

\paragraph*{Runtime Information (2.2)} Runtime values and types of the function's parameters and local variables during the failing test execution provide an LLM with concrete data about the program's behaviour.

\paragraph*{Angelic Forest (2.3)} For a given program location, a variable's \emph{angelic value}~\cite{chandra2011angelic} is a value that, if bound to the variable during the execution of a failing test, would enable the program to pass the test. Angelic forest~\cite{mechtaev2016angelix}, previously applied for synthesis-based repair, is a specification for a program fragment in the form of pairs of initial states and output angelic values, such that if the fragment satisfies these pairs, then the program passes the test. 

Inspired by this approach, we added a variant of angelic forest to a prompt; this variant combines variable values at the beginning of a function's execution coupled with the angelic values at the end of a function's execution, \ie the input/output requirements of the function. Since Python is dynamically typed, we specify both the values and types of variables. Angelic values can be computed using symbolic execution~\cite{mechtaev2016angelix,chandra2011angelic}; however, due to the immaturity of Python symbolic execution engines, we were unable to execute them on bugs in BugsInPy. Thus, we extracted angelic values from the correct versions of the programs via instrumentation.

\paragraph*{GitHub Issue (3.1)} A GitHub issue, when available, provides important contextual information for fixing the bug, as shown by Fakhoury~\etal~\cite{fakhoury2023towards}.

To denote subsets of $\mathbb{F}$, we utilise seven-width \emph{bitvectors}, where the $i$-th bit indicates whether the $i$-th fact in our taxonomy (\Cref{fig:facts}) is included in the set. For example, 0000100 corresponds to the set containing only the runtime information (2.2).

\subsection{Prompt Design}

We construct prompts via the prompt engineer $E \colon B \times 2^\mathbb{F} \rightarrow \Sigma^*$, which builds a prompt over the alphabet $\Sigma$ to repair an input bug using a subset of facts from $\mathbb{F}$. The prompt is constructed with the directive ``Please fix the buggy function provided below and output a corrected version" along with the included subset of facts. The buggy function's code, together with its docstring, is provided as part of the prompt for the LLM to effectively fulfill this directive. Each fact is incorporated via a specialised prompt template. \Cref{fig:aprprompt} shows an example prompt with incorporated facts, and the fact templates are detailed in supplementary materials (Appendix B). 

We employed the standard chain-of-thought~\cite{wei2022chain} approach by instructing the LLM to reason about the provided facts, as detailed in supplementary materials (Appendix C).

In our preliminary experiments, we discovered that LLM often generates incorrect import statements, which makes it hard to automatically extract patches from the responses and apply them to the code. To address it, we explicitly added the import statements in the current file to the prompts. A small study showed that this consistently improves the success rate, as detailed in supplementary materials (Appendix D). 

\section{The Fact Selection Problem}

Let an LLM be a function from a string, \ie a prompt, to a set of strings, the responses $R$. 
We consider an arbitrary measure $m: 2^R \times 2^C \rightarrow \mathbb{R}$ that scores a set of LLM responses \wrt some correctness criteria $C$ that maps correct patches to a high score and incorrect patches to a low score, and a prompt engineer $E \colon B \times 2^\mathbb{F} \rightarrow \Sigma^*$.

\begin{definition}[Fact Selection Problem]
Given a set of bug-relevant facts $\mathbb{F}$, a prompt engineer $E$, a buggy program $b$, and correctness criteria for that buggy program $C_b$, the 
 \emph{fact selection problem} is to find $F \subseteq \mathbb{F}$ that maximises 
\begin{equation}\label{eq:fact:selection}
  \arg\max_{F \in 2^\mathbb{F}} m(\text{LLM}(E(b, F)), C_b)
\end{equation}
\end{definition}
$C_b$ encompasses any of the standard correctness criteria such as a test suite or a specification, \etc
$\overset{*}{F}_b$ denotes an optimal solution of \Cref{eq:fact:selection} for $b$;
We use $\overset{*}{F}_b(\mathbb{F})$ to denote the optimal $F$ for the buggy program $b$ over the fact set $\mathbb{F}$.

Similarly, we use $\overset{*}F_B(\mathbb{F})$ to denote the optimal fact set $F$ over all the buggy programs $b \in B$ and the fact set $\mathbb{F}$. This can be defined as the solution to the equation below.
\begin{equation*}\label{eq:factset:selection}
   \arg\max_{F \in 2^\mathbb{F}}  \sum_{b \in B} m(\text{LLM}(E(b, F)), C_b)
\end{equation*}

We aim to answer the following questions in this section: 

\begin{itemize}

    \item \emph{How does the inclusion of each fact affect the overall effectiveness of a program repair prompt?}
    
    \item \emph{Is there point beyond which adding facts to a program repair prompt degrades its performance?}
        
    \item \emph{Can a fixed subset of facts be universally optimal up to a tolerance of $\epsilon$ for bug resolution across various bug sets?} 
    
\end{itemize}
The first question examines the impact of each fact on the repair effectiveness, questioning whether every fact contributes positively to the resolution process. \Cref{sec:eval:fact-utility} answers this question affirmatively, showing the inherent value of each fact. The second question delves into the potential for diminishing returns or even detrimental effects from overloading a prompt with too many facts, suggesting an optimal threshold for fact inclusion that maximises prompt efficacy. If there is no such point, then the optimal strategy will be to include all the facts. \Cref{sec:eval:monotonicity} shows that, on our dataset, adding facts to a prompt is non-monotonic. Formally, the final question asks whether, $\forall b \in B$, the following equation holds:
\begin{equation}
\label{eq:universal}
  m\big(\text{LLM}(E(b,\overset{*}{F}_B)),C_b\big) = m\big(\text{LLM}(E(b,\overset{*}{F}_b)), C_b\big) + \epsilon
\end{equation}
This equation asks whether one can select a fact set for a set of bugs that is as effective as a fact set tailored to each bug. \Cref{sec:eval:universal} answers this question by showing that this statement does not hold. The above answers, combined, establish the importance of the fact selection problem.

\subsection{Fact Utility}
\label{sec:eval:fact-utility}

A fact should only be considered for inclusion in a prompt if it has a potential to improve the outcome. To confirm the utility of the considered facts $\mathbb{F}$, we simplify the premise by assuming that facts are independent and pose two questions: ``What is the utility of each individual fact in improving repair performance on our dataset if we select the most effective fact set for each bug?'' and ``What is the utility of each individual fact in improving repair performance on our dataset if we select a random fact set for each bug?'' The first question addresses the potential effectiveness when we precisely know which facts to choose for a specific bug. This notion of utility, which we refer to as \emph{utility under optimal fact selection}, is relevant when we have a method to closely approximate an optimal solution $\overset{*}{F}_b(\mathbb{F})$ to \Cref{eq:fact:selection}, \ie choose the most effective facts for each bug $b \in B$. The second question explores the expected outcomes when we lack specific knowledge about which facts to select, and hence make a random choice. This notion of utility, which we call \emph{utility under uniform fact selection}, is relevant when solving \Cref{eq:fact:selection} is either difficult or infeasible. 

\begin{table}[t]
\centering

\resizebox{\linewidth}{!}{%
\begin{tabular}{lrrcrr}
\toprule
\multirow{2}{*}{\textbf{Fact}} & \multicolumn{2}{c}{\textbf{GPT-3.5}} & \multicolumn{2}{c}{\textbf{Llama3-70B}} \\
\cmidrule(lr{0.3em}){2-3}\cmidrule(lr{0.3em}){4-5}
 & \textbf{Gain} &  \multicolumn{1}{c}{\textbf{Shapley}} & \textbf{Gain} & \textbf{Shapley} \\
\midrule
Error Info.    & 0.48          & 0.54              & 0.41           & 0.32               \\
GitHub Issue   & 0.44          & 0.51              & 0.20           & 0.17               \\
Angelic Forest  & 0.08          & 0.11              & 0.36           & 0.28               \\
Failing Test   & 0.06          & 0.08              & 0.17           & 0.15               \\
Runtime Info. & 0.03          & 0.05              & 0.08           & 0.07               \\
Buggy Class   & -0.03         & -0.05             & 0.03           & 0.03               \\
Used Method S. & -0.12         & -0.18             & 0.01           & 0.00               \\

\bottomrule
\end{tabular}}
\caption{\changes{We report Gain (\Cref{eq:fact:gain}) and Shapley values (scaled by 16) for uniform fact selection on \BGPSetA. Gain quantifies the average percentage increase in prompt repair performance from the fact.}}
\vspace{-2mm}

\label{tab:fact_gain}
\end{table}

To estimate the utility of each fact, we generated prompts containing all subsets of the considered 7 facts (the remaining one, the buggy function, is always present in the prompt), which resulted in a total of 19228 prompts for \singleFunctionDataset, which is less than $314\times 2^7$ since some facts are not available for some bugs. For each prompt, we computed 15 responses to estimate the measures \passat{1} and \numfixed. This enabled us to both evaluate an optimal selection strategy by explicitly considering $\overset{*}{F}_b$ for each bug, and a uniform selection strategy.

We evaluate the utility of individual facts under uniform selection using two complementary measures: Shapley~\cite{winter2002shapley} and a new measure we introduce and call \emph{fact gain}, defined in \Cref{eq:fact:gain}. We report fact gain along with Shapley for two reasons:  (1) Shapley's results are hard to interpret and (2) we have the luxury of exhaustive enumeration, since our fact set is small. Fact gain computes the net increase in \passat{1} scores due to the addition of a specific fact $f$; we defined it by adapting relative change~\cite{relativeChange} to our problem domain by setting the reference value to fact subsets that do not contain the measured fact $f$.
Let $J_f = \{(b, F) \in J \mid f \in F\}$ be the subset of prompts whose fact set $F$ includes the fact $f$, and $J_{\overline{f}} = \{(b, F) \in J \mid f \notin F\}$ be those prompts that do not.  \changes{Let $R_f = \mathit{LLM}(J_f)$ and $R_{\overline{f}} = \mathit{LLM}(J_{\overline{f}})$}.  Then
the gain of each fact is:

\begin{equation}
\label{eq:fact:gain}
A(f) = \frac{\passat{k}(R_f) - \passat{k}(R_{\overline{f}})}{\passat{k}(J_{\overline{f}})}
\end{equation}
$A(f)$ computes the change in the likelihood of generating a successful repair when the fact $f$ can be used in a prompt.

\Cref{tab:fact_gain} showcases the significance of each fact in APR prompts, leveraging both aggregate gain ($A(f)$) and Shapley values, as defined in \Cref{eq:fact:gain}. 
Another interesting observation from the experimental results is that the facts ``Buggy Class'' and ``Used Method'' Signatures" exhibit a negative aggregate gain for GPT-3.5, indicating that their inclusion might adversely affect the repair outcome on average. 
\changes{
For Llama3-70B, these facts have an aggregate gain close to zero, still making them the least beneficial among the considered facts. Nonetheless, each of these facts allows GPT-3.5 to fix 4 additional bugs that were not fixed without them. For Llama3-70B, ``Buggy Class'' enables fixing 1 additional bug, while ``Used Method Signatures'' contributes to fixing 5 additional bugs. This shows the importance of fact selection.}

To demonstrate the utility of facts $\mathbb{F}$ under optimal fact selection, we compute the number of bugs that were fixed exclusively when each fact was available. \changes{These exclusive fixes are reported in the column ``\# Excl.'' for both GPT-3.5 and Llama3-70B in \Cref{tab:expected_bugs_fixed}. This table highlights the unique bug-fixing contribution of each fact, showing the distinct set of bugs resolved by each fact across the two models.
It further shows that all facts improve performance. These improvements are statistically significant for both models, with no p-value exceeding 0.004.
}

\begin{table}[t]
\centering
\resizebox{\linewidth}{!}{%
\begin{tabular}{lccccccc}
\toprule
\multirow{2}{*}{\textbf{Fact f}} & \multicolumn{3}{c}{\textbf{GPT-3.5}} & \multicolumn{3}{c}{\textbf{Llama3-70B}} \\
\cmidrule(lr{0.3em}){2-4}\cmidrule(lr{0.3em}){5-7}

                 & \textbf{\# Excl.} & $\overset{*}{F}_b(\mathbb{F} - \{f\})$ & $\mathbf{\Delta}$ & \textbf{\# Excl.} & $\overset{*}{F}_b(\mathbb{F} - \{f\})$ & $\mathbf{\Delta}$ \\
\midrule
Error Info.       & 9 & 0.331 & 0.071 & 8 & 0.257 & 0.0493 \\
Failing Test      & 7 & 0.375 & 0.027 & 7 & 0.281 & 0.0251 \\
Angelic Forest    & 7 & 0.371 & 0.032 & 5 & 0.245 & 0.0607 \\
Buggy Class       & 4 & 0.392 & 0.010 & 1 & 0.290 & 0.0166 \\
Used Method S.    & 4 & 0.393 & 0.009 & 5 & 0.285 & 0.0208 \\
GitHub Issue      & 3 & 0.341 & 0.062 & 6 & 0.263 & 0.0429 \\
Runtime Info.     & 2 & 0.384 & 0.019 & 8 & 0.288 & 0.0183 \\
\bottomrule
\end{tabular}}
\caption{\changes{\# Excl. shows the bugs that could only be fixed by including the specific fact under optimal fact selection \changes{for \BGPSetA}. $\overset{*}{F}_b(\mathbb{F} - \{f\})$ shows performance of the best facts when \textbf{f} cannot be selected. $\mathbf{\Delta}$ represents the drop in performance due to excluding the fact from the bug's best performing fact set.  
}}
\label{tab:expected_bugs_fixed}
\vspace{-2mm}

\end{table}

Second, we analysed each fact's utility under optimal selection by how its inclusion in or its exclusion affects \passat{k}. We do so by simulating the scenario where specific facts are missing: If a fact $f$ were missing, we would be forced to compute $\overset{*}{F}_b$ over $\mathbb{F} - \{f\}$ for each bug $b \in B$. The baseline for this scenario is when all the facts are available. 

Each row in the table in \Cref{tab:expected_bugs_fixed} details the \passat{1} attainable by the best prompts, with and without a specific fact (denoted by f). For each bug $b \in B$, prompts are constructed using optimal fact sets $\overset{*}{F}_b(\mathbb{F})$ over all facts and $\overset{*}{F}_b(\mathbb{F} - \{f\})$, excluding the fact $f$. The consistent reduction in \passat{1} (denoted as $\Delta$) emphasises the value of each fact in bug fixing.

To determine the statistical significance of the impact of each individual fact under optimal selection, we calculated \passat{1} scores for all bugs, both with and without a particular fact. These scores were then compared. the Wilcoxon signed-rank test shows that the inclusion of each fact has a statistically significant effect on each bug's optimal fact set.

\begin{framed}
\vspace{-2mm}
\begin{finding}\label{finding:utility}
\changes{Under the assumption that we select the most effective fact set for repairing each bug, each of the considered facts demonstrates its utility on our dataset.}
\end{finding}
\vspace{-2mm}
\end{framed}

When selecting an optimal fact set for repairing each bug, each of the seven considered facts proves useful on our dataset: including any of these facts in the prompt helps repair at least one bug exclusively and has a statistically significant positive impact on \passat{1}. 


\subsection{Impact of Fact Set Size on Prompt Performance}
\label{sec:eval:monotonicity}
In this section, we investigate the concept of prompt monotonicity by examining how the incremental addition of facts affects prompt performance. 
Monotonicity, in this context, refers to a consistent improvement in performance with each additional fact. This implies that more information invariably leads to better outcomes. Conversely, non-monotonicity indicates that there exists a threshold beyond which adding more facts does not enhance, and may even degrade, performance. 

Similarly to \Cref{sec:eval:fact-utility}, we evaluate the non-monotonicity of prompts in two settings: under an optimal fact selection and under a random selection. For each of them, we aggregate \passat{1} scores for all bugs over sets containing a varying number of facts. Figure~\ref{fig:non_monotonic} presents the performance of \passat{1} for the prompts across different fact set cardinalities, ranging from 0 to 7. The maximum \passat{1} corresponds to an optimal fact selection for each bug, and the average \passat{1} corresponds to a random fact selection.

\changes{From the plot, we observe that for both GPT-3.5 and Llama3-70B, the ``Max Pass@1'' scores (solid lines) generally increase with the number of facts, reaching a peak at 3 facts for GPT-3.5 (solid blue) and 4 facts for Llama3-70B (solid orange) before declining. The ``Avg Pass@1'' scores (dotted blue line) for GPT-3.5 show improvement with more facts, reaching a plateau between 2 and 5 facts, followed by a decrease. For Llama3-70B, the ``Avg Pass@1'' scores (dotted orange line) increase up to 4 facts and then decrease. This pattern confirms the non-monotonicity in prompt performance as the number of facts increases.}

\begin{framed}
\vspace{-2mm} 
\begin{finding}\label{finding:monotonic}
\changes{Prompt performance is non-monotonic with respect to the number of included facts. While adding facts generally improves performance, there exists a threshold beyond which additional facts hinder performance.}
\end{finding}
\vspace{-2mm}
\end{framed}


\subsection{Non-Existence of a Universally Optimal Fact Set}
\label{sec:eval:universal}
A "universally optimal fact set" in the context of automated program repair is a collection of facts that, when applied, yields the highest effectiveness in terms of bug fixes and pass@1 scores across a wide range of projects. For defining an universally optimal fact set, we first define the quality of a fact set in terms of the following properties:
 
\begin{itemize}
    \item \textbf{Efficiency:} A fact set is efficient when it outperforms alternative fact sets. 
    \item \textbf{Universality:} A fact set is universal when it is efficient up to $\epsilon$-tolerance, as defined in \Cref{eq:universal}.
    \item \textbf{Coverage:} The set of bugs a fact set can resolve.
\end{itemize}

We define the function $\mathit{Coverage}: 2^{\mathbb{F}} \rightarrow 2^B$ where $F \subseteq \mathbb{F}$ is a fact set and $B$ is the set of all bugs in the dataset being considered, such that $\mathit{Coverage}(F)$ returns the set of bugs fixed by the fact set $F$.

The Coverage Ratio for a given fact set $F$ is defined as:
\begin{equation}
    \mathit{CR}(F) = \frac{|\mathit{Coverage}(F)|}{|\bigcup_{\forall F_i \subseteq \mathbb{F}} \mathit{Coverage}(F_i) |}
\end{equation}
\noindent We prefer sets that maximise universality and coverage ratio.

\begin{table}[t]
\centering
\resizebox{\linewidth}{!}{%
\begin{tabular}{lcccccc}
\toprule
\multirow{2}{*}{\textbf{Project}} 
    & \multicolumn{3}{c}{\textbf{GPT-3.5}} & \multicolumn{3}{c}{\textbf{Llama3-70B}} \\
\cmidrule(lr{0.3em}){2-4}\cmidrule(lr{0.3em}){5-7}

 & \textbf{Best Fact Set} & \textbf{Project} & \textbf{Total} & \textbf{Best Fact Set} & \textbf{Project} & \textbf{Total} \\
\midrule
luigi           & 0011111 & 0.53 & 0.26 & 1010111 & 0.42 & 0.16 \\
black           & 0010111 & 0.38 & 0.22 & 1101011 & 0.35 & 0.17 \\
fastapi         & 0000101 & 0.19 & 0.18 & 0011001 & 0.11 & 0.11 \\
httpie          & 0001111 & 0.50 & 0.25 & 0001000 & 0.50 & 0.12 \\
pandas          & 0011001 & 0.25 & 0.25 & 0101011 & 0.13 & 0.17 \\
tornado         & 0011000 & 0.37 & 0.20 & 1001001 & 0.22 & 0.13 \\
ansible         & 0010101 & 0.10 & 0.12 & 0011011 & 0.15 & 0.14 \\
matplotlib      & 0001001 & 0.36 & 0.25 & 0101010 & 0.26 & 0.16 \\
cookiecutter    & 0001101 & 0.93 & 0.20 & 0001010 & 0.67 & 0.14 \\
tqdm            & 0001111 & 0.00 & 0.24 & 1011001 & 0.00 & 0.13 \\
youtube-dl      & 0011001 & 0.08 & 0.25 & 0001110 & 0.06 & 0.14 \\
keras           & 0101111 & 0.32 & 0.23 & 1011110 & 0.17 & 0.15 \\
scrapy          & 0011111 & 0.45 & 0.26 & 1101110 & 0.36 & 0.15 \\
sanic           & 0010101 & 0.64 & 0.21 & 0110011 & 0.51 & 0.15 \\
thefuck         & 0001001 & 0.16 & 0.21 & 0001001 & 0.29 & 0.12 \\
\bottomrule
\end{tabular}}
\caption{Comparison of project-specific best fact sets and their project's average Pass@1 scores and the total average Pass@1 scores across all projects \changes{in \BGPSetA}. The fact sets are represented using their bitvector encodings. \changes{The table highlights that different repositories have different best fact sets, indicating the importance of tailored fact selection for effective bug fixing.}}
\vspace{-3mm} 

\label{tab:project_performance}
\end{table}

\Cref{tab:project_performance} \changes{presents the best-performing fact sets for each project, along with their project-specific and overall \passat{1} scores for both GPT-3.5 and Llama3-70B. Notably, no single fact set achieves top performance across all projects. For instance, the fact set 0010101 achieves a \passat{1} of 0.21 on GPT-3.5 for the Sanic project, which is below the highest \passat{1} score of 0.26. Similarly, in the FastAPI project, the best fact set 0000101 attains a project-specific \passat{1} of 0.19 on GPT-3.5, slightly exceeding its overall score of 0.18, yet still lower than the highest score of 0.26 on the same model. A similar variability is observed with Llama3-70B, where no common fact set appears across different repositories, highlighting the unique nature of each project's optimal fact set. Additionally, the highest occurrence count for any fact set is just 2, with five fact sets appearing twice. This distribution emphasizes the inconsistency in fact set effectiveness across projects and suggests that a universally optimal fact set is unlikely, given the diverse nature of repositories and bugs.}

\begin{table}[t]
\centering

\begin{tabular}{lr}
\toprule
\textbf{Fact Aggregations}          & \textbf{\numfixed} \\
\midrule
Best Fact Set in \BGPSetA   & 77                               \\
Best Fact Set in \BGPSetB   & 84                               \\
All Facts   & 72                               \\
Bugs Fixed by the Top 5 Fact Subsets     & 99                               \\
Bugs Fixed by Any Fact Subset    & 119                            \\
\bottomrule
\end{tabular}
\caption{The number of bugs plausibly fixed by fact subset \changes{in \BGPSetB using GPT-3.5}.}
\label{tab:coverage}
\vspace{-3mm} 
\end{table}

\Cref{tab:coverage} assesses the effectiveness of various approaches of fact selection in producing plausible fixes on GPT-3.5. The analysis underscores the Best Fact Set, identified as $1111001$, which was selected for its highest coverage in terms of the number of bugs it could fix according to the training data, compared against broader approaches such as the Top 5 Union and Total Union. The Top 5 Union, which aggregates the bugs fixed by the top five best performing fact sets, generates fixes that pass tests for 99 bugs, while the Total Union, encompassing bugs fixed by all fact subsets, resolves 119 bugs. These unions significantly surpass the Best Fact Set in bug resolution capability, fixing many additional bugs (22 and 42 respectively). This shows that the highest coverage ratio of the fact set is $\mathit{CR}(1111001) = 0.65$ that it fixes 65\% of the bugs while missing $35\%$ of the bugs fixable by other sets.

These results highlight that the Best Fact Set does not have a high coverage ratio, especially compared to the theoretical maximum of an optimal fact selection which has a coverage ratio of 1. The coverage ratio's delineation as monotonic — in that the fact set with the highest number of bugs fixed is deemed the best — indicates that within this dataset, no fact set achieves a high coverage ratio.
\Cref{tab:coverage} further shows the limitations inherent in static fact selection strategies, as demonstrated by the performance of the Best Fact Sets \wrt \BGPSetB. These sets fix 84 bugs, and set the upper limit for universal fact selection in \BGPSetB. These sets were not ranked high in the training data, as they were positioned at ranks 16 and 34, respectively out of 128 candidates.

 \begin{figure*}[t]
 \centering
  \includesvg[inkscapelatex=false, width=0.9\linewidth]{pictures/fix_upset}
  \caption{\changes{This upset diagram compares the 5 fact sets that fix the most bugs in \BGPSetA with using no facts (encoded as $0000000$) including the buggy function and chain-of-thought instructions. 
  The fact sets shown on the left collectively 
  fix 107 bugs, 18 more than the single best fact set, which fixes 89 bugs. 
  The diagram is constructed by generating prompts from each fact set, sending them to GPT-3.5 for $n=15$ responses, and identifying the set of bugs with at least one passing response.}}
  \label{fig:upset}
    \vspace{-3mm} 

\end{figure*}

The UpSet diagram~\cite{lex2014upset} in \Cref{fig:upset} shows the combinatorial overlap among the top 5 fact sets from \BGPSetA, as well as a baseline which does not contain any facts, encoded with the bitvector $0000000$. The diagram is particularly instructive in revealing the number of bugs addressed by various intersections of these fact sets, with the largest subset intersection resolving 41 bugs. Each of the fact sets is shown to individually contribute to the resolution of up to 7 bugs. Cumulatively, these 6 fact sets fix a total of 107 bugs, surpassing the efficacy of the single best fact set, which fixes 89 bugs.
This UpSet plot helps us in answering the question of the existence of a universal fact set \wrt the number of bugs fixed, we would expect to see a row with dots in most, if not all, columns, signifying its presence in the majority of intersections. However, the absence of such a pattern in this upset plot indicates there is no single fact set that fixes all bugs. Instead, different fact sets are effective for different bugs.

\begin{framed}
\vspace{-2mm} 
\begin{finding}\label{finding:universal}
\changes{The diversity in the best fact sets across different repositories (\Cref{tab:project_performance}), and the significant difference in the bugs fixed by the top five fact sets (\Cref{fig:upset}), both point to the absence of a universal fact set in our dataset.}
\end{finding}
\vspace{-2mm}
\end{framed}

\subsection{Effect of Fact Order on Performance}

\changes{
The experiments conducted to this point assume a fixed fact order. We now investigate the effect of permuting the facts.
Our prompt template constrains the permutations we consider.
We add the facts "Buggy class" (1.1) and "Used method signatures" (1.2) to our prompt next to each other in the order that they appear in source code, so we do not separately permute them.
Similarly, the "Failing Tests" (1.3) fact immediately precedes the "Error Information" (2.1) fact it generates in the template. Thus, we  only permute 5 facts.}

\changes{
This experiment was conducted on \preliminaryStudyDataset, comprising 32 bugs. For each of the $120$ permutations, we created prompts for each bug, resulting in a total of $120 \times 32$ prompts. We computed the \passat{1} for each prompt, and then averaged these scores across the 32 bugs for each permutation. 
The resulting violet histogram in \Cref{fig:fact-order} represents the mean \passat{1} performance across these permutations. 
}

\changes{
We conducted a similar analysis for different fact subsets, also shown in \Cref{fig:fact-order}. With $7$ facts, there are $128 \times 32$ prompts. For each subset, we calculated the mean \passat{1} across the $32$ bugs. The yellow histogram  represents the distribution of the mean performance over these fact subsets. Comparing the two histograms reveals that the variability due to fact order (violet) is relatively narrow, ranging from $0.24$ to $0.31$. In contrast, the impact of different fact subsets (yellow) is significantly broader, ranging from $0.07$ to $0.34$. This demonstrates that the selection of which facts to include (fact subsets) has a much greater influence on prompt performance than the specific order in which these facts are presented.
}

 \begin{figure}[t]
 \centering
 \vspace{-2mm}

  \includesvg[inkscapelatex=false, width=\linewidth]{pictures/permutation_dist}
  \caption{\changes{Distribution of mean \passat{k} evaluated on \preliminaryStudyDataset. The violet histogram reports the mean performance under permutation; the yellow histogram reports mean by subset.
  The wider range in the yellow distribution suggests that fact subsets have a stronger impact on performance than fact order.}}
  \label{fig:fact-order}
  \vspace{-3mm} 

\end{figure}

\begin{framed}
\vspace{-3mm} 
\begin{finding}\label{finding:order}
\changes{Selecting facts has more impact on performance than ordering them.}
\end{finding}
\vspace{-2mm}
\end{framed}

\section{Selecting Facts with \projName}
\label{sec:lppr}

Universal fact selection, as demonstrated in \Cref{sec:eval:universal}, does not achieve consistent performance across all the bugs in our dataset. Thus, to automate creating bug-tailored prompts, we introduce \projName, a random forest trained to select relevant facts for inclusion in the prompts. 

We focus on the task of predicting the success or failure of test executions based on vectors representing features extracted from both the prompt and the code. 

Our training dataset $\mathcal{D}$ is constructed from the \BGPSetA. It consists of pairs $(j, y)$, where:
\begin{itemize}
    \item $j = (b, F) \in \mathbb{J}$ represents a job, which is a tuple consisting of a bug $b$ along with and a fact set $F \subseteq \mathbb{F}$
    \item \(y \in [0, 1]\) represents the probability of successfully fixing the bug $b$ in a single trial, given the fact set $F \subseteq \mathbb{F}$. This probability is computed using \passat{1}.
\end{itemize}
\changes{$\mathcal{D}$ consists of 157 bugs, each with 128 fact combinations, totaling $20,096$ samples. However, since not all facts are available for every bug, missing facts are labeled as ``None''. This reduces the number of unique prompts to $9496$.}

We manually craft a set of features $\mathbf{f}(j)$ based on domain knowledge and the characteristics of the facts. The feature function $\mathbf{f} : \mathbb{J} \rightarrow \mathbb{R}^m$ maps the input job $j \in \mathbb{J}$ to an $m$-dimensional feature space.

The goal is to train a machine learning model $\mathcal{M}$ capable of using the feature vector $\mathbf{f}(j)$ to accurately predict the likelihood of success. Specifically, the model is designed to learn a function $\mathcal{M}: \mathbb{R}^m \rightarrow [0,1]$, aiming to optimise the accuracy of success predictions as follows:
\begin{equation}
    \max_{\mathcal{M}} \; \big(\mathcal{M}(\mathbf{f}(j)) = y \;|\; J, y \in \mathcal{D}\big).
\end{equation}
This is achieved through an appropriate training process that adjusts the parameters of $\mathcal{M}$ based on the training data $\mathcal{D}$.

\subsection{Feature Selection}
\label{sec:maniple:select}
To train a machine learning model to meet these objectives, we define the feature vector, $\mathbf{f}(\mathbf{x}) = [\mathbf{b}, \mathit{rep\_id}, \ell, c]^T$, where
\begin{itemize}
    \item Bitvector ($\mathbf{b}$): encodes the fact set $F$, where $\mathbf{b} \in \{0, 1\}^n$.
    \item Repository ID ($\mathrm{rep\_id}$): uniquely identifies the source repository of the bug $b$.
    \item Prompt Length ($\ell$): the length of the prompt in either characters or tokens, where $\ell \in \mathbb{N}^{0}$. Cross-validation determines whether characters or tokens are chosen.
    \item Cyclomatic Complexity ($c$): a measure of code complexity that quantifies the number of linearly independent paths through a program's source code.
\end{itemize}

This choice of features was guided by the investigation conducted on \BGPSetA. Specifically, the including Repository ID ($\mathit{rep\_id}$) is supported by the findings in \Cref{tab:project_performance}, which illustrate significant variability in the optimal bitvectors for fact selection across different projects. This variability underscores the influence of the repository context on successful repair. Additionally, Prompt Length ($\ell$) was identified as a crucial factor, displaying a Spearman correlation of $-0.18$ with the \passat{1} for repair success, accompanied by a highly significant p-value of $p < 10^{-129}$. This correlation holds for both token and character lengths of prompts, indicating that an increase in prompt length is associated with a decrease in LLM performance --- a conclusion that is further supported by the non-monotonicity of adding facts to prompts, as noted in \Cref{fig:non_monotonic}, as more facts increase prompt length.

Cyclomatic Complexity emerged as another pivotal feature, showing a negative Spearman correlation of $-0.1$ with the \passat{1}, p-value of $10^{-42}$. Unlike prompt length, Cyclomatic complexity does not depend on the fact set chosen (recall that the buggy code itself is necessarily always included) but remains instrumental in predicting the \passat{1} for a bug, mainly for scenarios where no prompt generates a successful fix.

\subsection{\projName: A Random Forest for Fact Set Selection}

Leveraging these observations, we present \projName, a random forest model for the fact selection task and evaluate it in both regression and classification settings. As a regressor, the model directly predicts the \passat{1} for each job. In the classification task, we categorise the scale (i.e., [0, 1]) based on the number of fact sets considered by the model. Our analysis reveals that classification performs better.
This finding can be attributed to the noise and significant variance present in our data, as detailed in supplementary materials (Appendix A). 
The classification method proved more robust to variance in \passat{1} compared to regression. Additionally, ordering and ranking the fact sets did not yield comparable performance. This is likely due to the variance in ranks, which directly stems from the variance in \passat{1}. %

Additionally, we trained \projName only on the top five highest-performing fact sets. These sets were identified using bootstrap aggregation, where we evaluated the performance of each fact set across multiple bootstrap samples drawn from our dataset. By aggregating the outcomes, we identified the best performing fact sets.
We optimised the model's hyperparameters through a comprehensive grid search, evaluating the performance of each combination of parameters. To ensure the model's generalisability and to prevent overfitting, we employed $k$-fold cross-validation with $k = 5$.

\section{Comparing \projName with SOTA LLM-Based APR}
\label{sec:maniple_eval}

We compared \projName with existing zero-shot non-conversational LLM-based APR methods that incorporate various types of information into prompts. Our baselines include approaches whose prompts include the following facts:

\begin{itemize}
    \item Buggy Function only, denoted as $T_0$
    \item Buggy Function, Buggy Class and Used Method Signatures, denoted as $T_1$, an approach similar to the technique by Chen~\etal~\cite{chen2024large}.
   \item Buggy Function combined with GitHub Issue, denoted as $T_2$, an approach similar to the technique by Fakhoury~\cite{fakhoury2023towards}.
    \item Buggy Function alongside Error Information, denoted as $T_3$, an approach similar to the technique by Keller~\etal~\cite{Keller2024AI}.
\end{itemize}

For a fair comparison, we supply the same prompts, built from these facts, to \projName and all the baselines.
These prompts, unsurprisingly, differ structurally from the prompts on which the baselines were run.  For example, our prompt incorporates our chain-of-thought instructions defined in supplementary materials (Appendix C). 
Our focus here is on comparing \projName's performance to the baselines' \wrt facts, not their performance given approach-specific optimal prompts, whose construction would require close cooperation with the authors of each baseline. %


We evaluated tool efficacy according to two criteria, \changes{based on $n=15$ responses (\Cref{sec:setup})}: (1) number of plausible fixes (\ie \cref{eq:numfixed}), and (2) the number of correct fixes, determined by the first plausible patch generated by the LLM and subsequently subjected to manual evaluation.

For assessing patch correctness, we employ an interrater agreement scale, where ``3'' indicates patches syntactically equivalent to the developer's patch, allowing for minor refactoring or restructuring, ``2'' --- patches that achieve the intended outcome through an alternate method, ``1'' --- diverging from the developer's patch, rendering correctness indeterminable, and ``0'' --- incorrect patches (irrelevant, incomplete, introducing regressions).
Each patch undergoes evaluation by two independent raters. In cases of scoring discrepancies, the raters discuss them. If the discrepancy is not resolved, the more conservative (lower) score is recorded, ensuring a rigorous standard of correctness. Following this scoring system, patches with the scores of 2 and 3 are labelled as "correct", and with the scores of 1 and 0 are labelled as "incorrect."

\begin{table}[t]
\centering
\begin{tabular}{lrrr}
\hline
\textbf{Tool}          & \textbf{\numfixed}   & \textbf{Correct} & \textbf{\% Correct}  \\
\hline
$T_0$  &  44 & 7 & 16\%  \\
$T_1$ & 37  & 5 & 14\%  \\
$T_2$  & 63 &  18  & 29\% \\
$T_3$  &  66 &  31 & 47\% \\
\projName & 88 & 37 & 42\% \\
\hline
\end{tabular}
\caption{\changes{Comparison of tool performance on the \BGPSetB test set, focusing on bugs fixed with GPT-3.5}. \textbf{\numfixed} represents the number of bugs with test-passing patches, and \textbf{Correct} indicates bugs fixed with patches identical to the developer's. We observe that \projName outperforms all the fact set combinations used. This shows that bug-tailored fact selection can improve the repair success.}
  \vspace{-3mm} 

\label{tab:tool_performance_comparison}
\end{table}

The analysis in \Cref{tab:tool_performance_comparison} underscores the potential of bug-tailored prompt selection for boosting automated program repair efficacy. Previous LLM-based zero-shot non-conversational approaches, represented by $T_1$, $T_2$ and $T_3$, that always use the same set of facts achieve a maximum of 66 plausibly fixed patches on \BGPSetB. In contrast, \projName, which leverages bug-tailored fact selection, identifies a significantly higher number of plausible patches (88). Furthermore, \projName boasts 6 additional correct fixes compared to the best of these approaches. These findings suggest that tailoring the fact set to the specific context of each bug has the potential to improve the upper bound for repair success.

Surprisingly, $T_1$, which utilises function code alongside scope information (including class information and invoked functions), fixes fewer bugs compared to using function code alone. However, this does not imply that scope and class information are useless. $T_1$ still identifies 5 plausible patches, 2 of which are correct, that would not be found using just the function code as the only fact.

\section{Threats to validity}
\label{sec:threats}

\changes{In assessing LLMs, we recognise the challenge of potential data leakage, as some training datasets are not publicly available. This limits our ability to know exactly what information the model has encountered. However, our primary focus is on the impact of different facts on repair performance, rather than the performance itself. For instance, if a prompt with an ``angelic forest'' leads to a bug fix while a prompt without it does not, this highlights the significant role of ``angelic forest'' in enhancing bug-fixing performance. This remains true regardless of the bug's presence in the training data. Thus, our findings are robust even with possible data leakage, and using the same model across our baselines partially mitigates this issue.}


The facts collected for our study represent our best effort. Although, to the best of our knowledge, this work considers the widest range of information in APR literature, we acknowledge that our fact set is provisional and will undoubtedly change as research into LLM-assisted APR continues.

The external validity of our findings relies on the distribution of bugs within our dataset. Our dataset is 
a subset of BugsInPy, a benchmark published at FSE'20~\cite{widyasari2020bugsinpy}.  BugsInPy is a curated dataset built to best practice from GitHub repos with more than 10k stars that have at least one test case in the fixed commit that distinguishes the buggy version from its fix.  Its representativeness has not been challenged and, from first principles, we can think of no reason that our filtering (\Cref{sec:back:bench}) would introduce systematic bias. 

\section{Related Work}
\label{sec:related}

This work is relevant to the areas of prompt engineering and automated program repair.

\paragraph*{Prompt Engineering} 
Recent advancements in prompt engineering have significantly influenced the effectiveness of models like ChatGPT. Notably, the tree-of-thoughts approach~\cite{yao2023tree} and the zero-shot-CoT approach~\cite{kojima2022large} have emerged as pivotal strategies. Frameworks like ReACT~\cite{yao2022react} use LLM to generate reasoning traces and task specific actions in an interleaved manner. Self Consistency~\cite{wang2022self} is an approach that traverses multiple diverse reasoning paths through few shot CoT and uses the generations to select the most consistent answer. Automatic Prompt Engineer~\cite{zhou2022large} proposes a framework for automated prompt generation. It frames the task as a natural language synthesis task to construct prompts. This is orthogonal to our approach, as our task is to select ideal facts and the task of the Automatic Prompt Engineer is to refine the prompt into which the selected facts can be directly plugged. Repository Level Prompt Generation (RLPG)~\cite{shrivastava2023repository} is a very general framework for retrieving relevant repository context and constructing prompts, instantiated for code completion. RLPG generates prompt proposals and uses a classifier to choose the best one. In contrast, our fact selection problem aims to find an optimal combination of facts to repair a given bug. Our work also uses a wider variety of information, incorporating, apart from code context, dynamic and external information.

\paragraph*{Traditional Program Repair} Traditional APR techniques use search, \eg GenProg~\cite{le2011genprog}, or program synthesis such as SemFix~\cite{nguyen2013semfix}, Angelix~\cite{mechtaev2016angelix}, SE-ESOC~\cite{mechtaev2018symbolic}, and Trident~\cite{parasaram2021trident}. This study borrows the concept of angelic forest~\cite{mechtaev2016angelix} from the synthesis-based tools as one of the considered facts, demonstrating its utility in LLM-based APR.

\paragraph*{Program Repair with Contextual Information} CoCoNut~\cite{lutellier2020coconut} utilizes surrounding contextual information to train an ensemble of neural machine translation models. Rete~\cite{parasaram2023rete} employs Conditional Def-Use chains as context for CodeBert~\cite{feng-etal-2020-codebert}. CapGen~\cite{wen2018context} utilises AST node information to estimate the likelihood of patches. DLFix~\cite{li2020dlfix} treats the program repair task as a code transformation task, learning to transform by additionally incorporating the surrounding context of the bug. The context used in this work includes the class and scope information of the buggy program, which is broader than the contexts used above. FitRepair~\cite{xia2023revisiting} constructs prompts using identifier extracted from lines that look similar to the buggy line. Although, our work does not directly provide identifiers statically, as python is dynamic, we provide dynamic values of the variables during the test run.


\paragraph*{LLM-based Program Repair} LLM based techniques are making strides in APR.
InferFix~\cite{jin2023inferfix} uses few-shot prompting to repair issues from Infer static analyser. ChatRepair~\cite{xia2023keep} uses interactive prompting constructed using failing test names and their corresponding failing assertions.  Our approach focuses on the zero-shot, non-conversational setting, but can be potentially integrated with InferFix and ChatRepair. Various approaches focused on program engineering for APR, \eg incorporating bug-related information within the prompts~\cite{jin2023inferfix,xia2022practical,jiang2023impact}, as the quality of fixes could be enhanced by integrating contextual information, such as the bug's local context~\cite{prenner2024out} and details about relevant identifiers~\cite{xia2023revisiting}, into the prompt. Xia~\etal~\cite{xia2023revisiting} utilize relevant identifiers to augment the fix rate of prompts. Keller~\etal~\cite{Keller2024AI} reveals that, for debugging tasks, focusing on the specific line indicated by a stack trace is more effective than providing the entire trace. Similarly, Fakhoury~\etal~\cite{fakhoury2023towards} investigates how combining issue descriptions with bug report titles and descriptions enhances program repair efforts. Following recent research~\cite{ahmed2023improving}, we refer to such pieces of information as \emph{facts}.
Our work uses individual facts from previous work to formulate and motivate the fact selection problem.

\paragraph*{Program Repair for Python}
QuixBugs~\cite{lin2017quixbugs} is a benchmark consisting of small programs in Java, and Python. They are not reflective of real software projects. Bugswarm~\cite{tomassi2019bugswarm}, was constructed by automatically mining failing CI builds, and thus contains issues outside of the scope of our study, such as configuration issues.
BugsInPy~\cite{widyasari2020bugsinpy} manually curates 501 bugs from 17 popular Python Projects. We selected 314 bugs from this benchmark that require modifications within a single function, and which we managed to reproduce. Rete~\cite{parasaram2023rete}, a program repair tool that leverages contextual information, was evaluated on both Python and C, using BugsInPy as its Python benchmark. We did not compare \projName with Rete, because it relied on the line-granular perfect fault localisation (PFL), while this study uses function granular PFL. PyTER~\cite{oh2022pyter} is a program repair technique that focuses on Python TypeErrors; it was evaluated on a custom benchmark for type errors. 



\section{Conclusion}
\label{sec:conclusion}
In this paper, we explore the construction of effective prompts for LLM-based APR, leveraging facts extracted from the buggy program and external sources. Through a systematic investigation, we incorporate seven bug-relevant facts into the prompts, notably including angelic values, a factor previously not considered in this domain. Furthermore, we define the fact selection problem and demonstrate that a universally optimal set of facts for addressing various bugs does not exist. Building on this insight, we devise a bug-tailored fact selection strategy enhancing the effectiveness of APR.

\bibliographystyle{IEEEtran}
\bibliography{bibliography}
\appendices

\section{Analyzing the impact of nondeterminism on LLM's performance}
\label{sec:appendix:nondet}
In this investigation, we aim to determine the influence of nondeterminism on the performance of Large Language Models (LLMs). The study is conducted using \preliminaryStudyDataset, as it is expensive to run a query a large amount of responses from the LLM. A challenge presented by nondeterminism in LLMs is the variability in outcomes from one experiment to the next, precluding the use of the conventional performance evaluation method of generating $k$ responses from the LLM and determining success if at least one response meets the success criteria. However, due to the variability inherent in each trial, this approach may yield inconsistent results.

To address this variability, we use pass@k as our measure. Pass@k represents the probability of achieving at least one successful outcome within $k$ attempts at solving a problem. This is determined by soliciting $n > k$ responses from the API and calculating the likelihood of at least one success within $k$ trials, as discussed in \Cref{sec:background:nondet}. 

For our methodology, we obtained $n=30$ responses and evaluated pass@k for $k=1$. To simulate multiple runs, we employed bootstrapping by sampling $n$ responses from the pool of 30 responses with replacement, repeated $10$ times. The standard deviation of pass@1 for these $10$ samples was computed. Furthermore, we calculated the mean standard deviation of all the bugs in \preliminaryStudyDataset.

As depicted in \Cref{fig:nondet}, the mean standard deviation of pass@1 demonstrates diminishing returns from approximately $n=15$. This observation prompted our decision to choose $n=15$.
\Cref{fig:heatmap} displays the mean standard deviation of pass@k across varying response counts ($n$) and trial counts ($k$). Notably, the heatmap indicates a decrease in standard deviation with increasing $n$, while an increase in $k$ corresponds to higher standard deviation.

In understanding the mean standard deviation of pass@k, it is crucial to consider the granularity of the measure. Granularity refers to the smallest increment by which the measure can vary, encompassing all possible values within its range. Lower granularity allows for the capture of finer differences. However, if the mean standard deviation significantly exceeds the granularity, it can suggest that much of the finer differences in the output could be attributed to noise stemming from the nondeterminism of the LLM's output.

For our analysis, we specifically focus on $k=1$ because the granularity for pass@1 is $\frac{1}{n}$ for $n$ responses. 
This granularity allows for a more precise measurement of success probability since the standard deviation is lower than the granularity. 
However, for $k=2$, the granularity for pass@2 is $\frac{1}{\binom{n}{2}}$, which is relatively small compared to any chosen $n$ from our current data. 
For instance, pass@2 for $n=15$ yields $0.05$, which is larger than $\frac{1}{\binom{15}{2}} \approx 0.01$. Hence, we adopt $k=1$ to ensure more accurate results.

\begin{figure}[t]
 \centering
  \includesvg[inkscapelatex=false, width =\linewidth]{pictures/nondet}
  \caption{This plot illustrates the relationship between the standard deviation of pass@1 and the number of responses ($n$). The reduction in standard deviation demonstrates diminishing returns as $n$ increases. This observation motivated our selection of $n=15$, as its standard deviation is comparable to that of $n=30$.}
  \label{fig:nondet}
\end{figure}

\begin{figure}[t]
 \centering
  \includesvg[inkscapelatex=false, width =\linewidth]{pictures/heatmap}
  \caption{This heatmap depicts the standard deviation of pass@k against varying response count ($n$) and trial counts ($k$). The vertical gradient transitions from lighter to darker shades as $n$ increases, signifying a reduction in standard deviation and thereby highlighting enhanced measurement precision with number of queries. On the horizontal axis, the gradient shifts from darker to lighter shades as the $k$ grows, indicating an increase in standard deviation. This pattern suggests that lower values of $k$ and higher values of $n$ are associated with more precise outcomes.}
  \label{fig:heatmap}
\end{figure}

\section{Fact Prompt Templates}
\label{sec:appendix:fact_templates}

This section defines fact templates that are used to generate APR prompts.

\paragraph*{Buggy Function (Directive)} The body of the buggy function is given as the first section using the following template:
\begin{lstlisting}[escapechar=?,breaklines=true,frame=single]
Please fix the buggy function provided below and output a corrected version.
Following these steps:
?\hl{<CHAIN-OF-THOUGHT INSTRUCTIONS>}?

# The source code of the buggy function
```python
# this is the buggy function you need to fix
?\hl{<FUNCTION BODY>}?
```
\end{lstlisting}

\paragraph*{Buggy Class (1.1)} The declaration of a class containing the buggy function is added to the function body section:
\begin{lstlisting}[escapechar=?,breaklines=true,frame=single]
# The source code of the buggy function  
```python
# The declaration of the class containing the buggy function
  class ?\hl{<CLASS DECLARATION>}?:
  ?\hl{...}?  

  # this is the buggy function you need to fix
  ?\hl{<FUNCTION BODY>}?
```
\end{lstlisting}
A class docstring offers insights into the overall purpose and functionality of the class, which can guide the LLM in understanding how the buggy function should operate. It is added to the buggy class declaration in the prompt using the standard Python docstring notation.

\paragraph*{Used Method Signatures (1.2)} The methods used within the buggy function are incorporated into the buggy class declaration section: 
\begin{lstlisting}[escapechar=?,breaklines=true,frame=single]
The source code of the buggy function
```python
# The declaration of the class containing the buggy function
class ?\hl{<CLASS DECLARATION>}?:
  ?\hl{...}?

  # This function from the same class is called by the buggy function
  def ?\hl{<FUNCTION SIGNATURE>}?:
    # Please ignore the body of this function

  ?\hl{...}?
```
\end{lstlisting}

Signatures of the methods used, which are declared outside the class of the buggy function, are incorporated into the prompt as follows:
\begin{lstlisting}[escapechar=?,breaklines=true,frame=single]
# Buggy function source code
```python
# This function from the same file, but not the same class, is called by the buggy function
def ?\hl{<FUNCTION SIGNATURE>}?:
    # Please ignore the body of this function
  
  ?\hl{...}?
```
\end{lstlisting}

\paragraph*{Failing test (1.3)} The test code is incorporated into prompts in a separate section:

\begin{lstlisting}[escapechar=?,breaklines=true,frame=single]
# A test function that the buggy function fails:
```python
# The relative path of the failing test file: ?\hl{<TEST FILE NAME>}?

?\hl{<TEST CODE>}?
```
\end{lstlisting}

\paragraph*{Error Information (2.1)} We incorporate the error message and the stack trace into a separate section of the prompt:
\begin{lstlisting}[escapechar=?,breaklines=true,frame=single]
# The error message from the failing test
```text
?\hl{<ERROR MESSAGE>}?

?\hl{<STACK TRACE>}?
```
\end{lstlisting}

\paragraph*{Runtime Information (2.2)} Assume that $x_1$, ..., $x_n$ are local variables in the buggy function, $v_1$, ..., $v_n$ and $t_1$, ..., $t_n$ are their values and types at the beginning of the function's execution, and $v'_1$, ..., $v'_n$ and $t'_1$, ..., $t'_n$ are their values and types at the end of the function's execution. To represent runtime values inside the prompt, we use the following format.

\begin{lstlisting}[escapechar=?,breaklines=true,frame=single]
# Runtime values and types of variables inside the buggy function

Each case below includes input parameter values and types, and the values and types of relevant variables at the function's return, derived from executing failing tests. If an input parameter is not reflected in the output, it is assumed to remain unchanged. Note that some of these values at the function's return might be incorrect. Analyze these cases to identify why the tests are failing to effectively fix the bug.

# Case ?\hl{<CASE ID>}?
## Runtime values and types of the input parameters of the buggy function
?\hl{$x_1$}?, value: ?\hl{$v_1$}?, type: ?\hl{$t_1$}?
?\hl{...}?  
?\hl{$x_n$}?, value: ?\hl{$v_n$}?, type: ?\hl{$t_n$}?

## Runtime values and types of variables right before the buggy function's return
?\hl{$x_1$}?, value: ?\hl{$v'_1$}?, type: ?\hl{$t'_1$}?
?\hl{...}?  
?\hl{$x_n$}?, value: ?\hl{$v'_n$}?, type: ?\hl{$t'_n$}?

?\hl{...}?
\end{lstlisting} 

\paragraph*{Angelic Values (2.3)} Assume that the function operates a set of variables $x_1$, ..., $x_n$, for which the runtime values in the beginning of the function are $v_1$, ..., $v_n$, and the types are $t_1$, ..., $t_n$, and the angelic values at the end of the function are $a_1$, ..., $a_n$ and the types are $\mathit{at}_1$, ..., $\mathit{at}_n$. Then, this information is incorporated into the prompt as follows:

\begin{lstlisting}[escapechar=?,breaklines=true,frame=single]
# Expected values and types of variables during the failing test execution
Each case below includes input parameter values and types, and the expected values and types of relevant variables at the function's return. If an input parameter is not reflected in the output, it is assumed to remain unchanged. A corrected function must satisfy all these cases.

# Expected case ?\hl{<CASE ID>}?
# The values and types of buggy function's parameters
?\hl{$x_1$}?, expected value: ?\hl{$v_1$}?, type: ?\hl{$t_1$}?
?\hl{...}?  
?\hl{$x_n$}?, expected value: ?\hl{$v_n$}?, type: ?\hl{$t_n$}?

## Expected values and types of variables right before the buggy function's return
?\hl{$x_1$}?, expected value: ?\hl{$a_1$}?, type: ?\hl{$\mathit{at}_1$}?
?\hl{...}?  
?\hl{$x_n$}?, expected value: ?\hl{$a_n$}?, type: ?\hl{$\mathit{at}_n$}?

?\hl{...}?
\end{lstlisting} 
To reduce the length of the prompt, we only print the values of variables that change after the function execution. Values converted into a human-readable representation using Python's \texttt{.\_\_str\_\_()} method. Also, class and function variables are filtered.

\paragraph*{GitHub Issue (3.1)} We incorporate the GitHub issue's title and description into the prompt as follows:
\begin{lstlisting}[escapechar=?,breaklines=true,frame=single]
# A GitHub issue for this bug
The issue's title:
```text
?\hl{<ISSUE TITLE>}?
```
```text
The issue's detailed description:
?\hl{<ISSUE DESCRIPTION>}?
```
\end{lstlisting} 

\section{Chain-of-thought Instructions}
\label{sec:appendix:cot}

When writing these prompts, we also applied the standard techniques, chain-of-thoughts (CoT) prompting. Since we considered different possible subsets of facts for the inclusion in APR prompts, we used the following instruction template at the beginning of each prompt:
\begin{lstlisting}[escapechar=?,breaklines=true,frame=single]
  1. Analyze the failing test case and its relationship with ?\hl{<LIST\_OF\_FACTS>}?.
  2. Identify the potential error location within the problematic 
function.
  3. Explain the bug's cause using:
        ?\hl{<LIST\_OF\_FACTS>}?
  4. Suggest possible approaches for fixing the bug.
  5. Present the corrected code for the problematic function such that it satisfied the following:
        ?\hl{<CORRECTNESS\_CRITERIA>}?
\end{lstlisting}
For $\mathrm{bitvector} = 1111111$, the following will be its chain of thought instruction:
\begin{lstlisting}[escapechar=?,breaklines=true,frame=single]
  1. Analyze the failing test case and its relationship with the error message along with the buggy function, buggy class, buggy file, the github issue, the expected and actual input/output variable information .
  2. Identify the potential error location within the problematic 
function.
  3. Explain the bug's cause using:
        a. The buggy function
        b. The buggy class
        c. The buggy file
        d. The failing test and error message
        e. Discrepancies between expected and actual input/output variable values 
        f. The Github Issue information
  4. Suggest possible approaches for fixing the bug.
  5. Present the corrected code for the problematic function such that it satisfied the following:
        a. Passes the failing test.
        b. Satisfies the expected input/output variable values provided.
        c. Successfully resolves the issue posted in Github 
 \end{lstlisting}

We conducted a small-scale experiment on BGP32 to confirm the effectiveness of CoT. The proportion of prompts that successfully led to a fix, incorporating CoT, stood at 0.46; this figure fell to 0.38 in the absence of CoT. Our findings indicated a modest positive Spearman correlation of $0.08$ between the application of CoT and the pass@1 rate for repair success, which was supported by a statistically significant p-value of $10^{-13}$. This led us to directly include Chain of Thought into our prompts.

\section{Handling Imports in Prompts}
\label{sec:appendix:imports}

Without listing import statements of the current file in APR prompts, the LLM may generate arbitrary import statements and call functions that are not defined within the repository. This behavior can be attributed to two main reasons. First, the LLM might lack contextual knowledge about the identifiers used within the buggy function. Second, it might assume the absence of import statements in the program and, based on its training data, add imports. Note that this is often not related to the functional correctness of generated code, but incorrect imports make it hard to extract patches from the LLM's responses and insert them into the buggy programs.

For example, consider pandas:84. The LLM encounters a piece of code that uses the \texttt{BlockManager} identifier, which is an internal component of `pandas`. Without specific context or import statements, the LLM might incorrectly suggest importing \texttt{BlockManager} directly from the top-level \texttt{pandas} package with:
\begin{lstlisting}[escapechar=?,breaklines=true]
from pandas import BlockManager
\end{lstlisting}
However, the appropriate way to import \texttt{BlockManager} is from within the \texttt{pandas.core.internals} module, which is more specific and not immediately apparent without domain knowledge or explicit instruction:
\begin{lstlisting}[escapechar=?,breaklines=true]
from pandas.core.internals import BlockManager
\end{lstlisting}
To test whether adding import statements improves pass@k, we manually selected 10 bugs which frequently resulted in undefined identifier errors and conducted two experiments for a comparative study. In the first experiment, we enumerated all 64 possible bitvectors, set the seed to 42, and the temperature to 1, and obtained the fix patches from the LLM. The pass@5 calculated in this experiment was 0.233. Then, we kept the settings the same and added the following instruction at the beginning of our prompt to obtain fix patches from the LLM:

\begin{lstlisting}[escapechar=?,breaklines=true,frame=single]
Assume that the following list of imports is available in the current environment, so you do not need to import them when generating a fix.
```python
<import statements>
```
\end{lstlisting}

The pass@5 calculated from the above example was 0.271, which suggests a noticeable enhancement in the fix rate.

We did not consider import statements as a fact for our study, because it mostly solved a technical issue that helped us extract patches from the responses, rather than affecting the functional correctness of the generated code.

\end{document}